\documentclass[aps,prx,article,twocolumn,preprintnumbers,amsmath,amssymb,superscriptaddress]{revtex4}

\date{\today}
\usepackage{epsfig}
\usepackage{cases}
\usepackage{subfigure}
\usepackage{amsmath}
\usepackage{graphicx}
\usepackage{dcolumn}
\usepackage{bm}
\usepackage[colorlinks,linkcolor=blue,hyperindex,CJKbookmarks]{hyperref}
\usepackage{color}
\usepackage{float}
\usepackage{mathtools}
\usepackage{hyperref}
\usepackage{comment}
\usepackage{bbm}
\begin{document}
\newcommand{\Ham}{\mathcal{H}}
\newcommand{\kbf}{\mathbf{k}}
\newcommand{\qbf}{\mathbf{q}}
\newcommand{\Qbf}      {\textbf{Q}}
\newcommand{\lbf}      {\textbf{l}}
\newcommand{\ibf}      {\textbf{i}}
\newcommand{\jbf}      {\textbf{j}}
\newcommand{\rbf}      {\textbf{r}}
\newcommand{\Rbf}      {\textbf{R}}
\newcommand{\Schrdg} {{Schr\"{o}dinger}}
\newcommand{\aband} {{(\alpha)}}
\newcommand{\bband} {{(\beta)}}

\newcommand{\eps} {{\bm{\varepsilon}}}
\newcommand{\probA}      {{\mathsf{A}}}
\newcommand{\HamPump}      {{\Ham_{\rm pump}}}
\newcommand{\HamPr}      {{\Ham_{\rm probe}}}
\newcommand{\timeMax} {{t_{\rm m}}}
\newcommand{\qin} {{\qbf_{\rm i}}}
\newcommand{\qout} {{\qbf_{\rm s}}}
\newcommand{\epsin} {{\eps_{\rm i}}}
\newcommand{\epsout} {{\eps_{\rm s}}}
\newcommand{\win} {{\omega_{\rm in}}}
\newcommand{\wout} {{\omega_{\rm s}}}

\setlength{\parindent}{2ex}

\title{X-Ray Scattering from Light-Driven Spin Fluctuations in a Doped Mott Insulator}

\author{Yao Wang}
 \email[All correspondence should be addressed to Y.W.(\href{mailto:yaowang@g.clemson.edu}{yaowang@g.clemson.edu}) and M.M. (\href{mailto:mmitrano@g.harvard.edu}{mmitrano@g.harvard.edu}) 
]{}
 \affiliation{Department of Physics and Astronomy, Clemson University, Clemson, South Carolina 29631, USA}
\author{Yuan Chen}
  \affiliation{Department of Applied Physics, Stanford University, Stanford, California 94305, USA}
\affiliation{Stanford Institute for Materials and Energy Sciences, SLAC National Accelerator Laboratory, 2575 Sand Hill Road, Menlo Park, California 94025, USA.}
\author{Thomas P. Devereaux}
\affiliation{Stanford Institute for Materials and Energy Sciences, SLAC National Accelerator Laboratory, 2575 Sand Hill Road, Menlo Park, California 94025, USA.}
\affiliation{Department of Materials Science and Engineering, Stanford University, Stanford, California 94305, USA.}
\affiliation{Geballe Laboratory for Advanced Materials, Stanford University, Stanford, California 94305, USA.}
\author{Brian Moritz}
\affiliation{Stanford Institute for Materials and Energy Sciences, SLAC National Accelerator Laboratory, 2575 Sand Hill Road, Menlo Park, California 94025, USA.}
\author{Matteo Mitrano}
 \email[All correspondence should be addressed to Y.W.(\href{mailto:yaowang@g.clemson.edu}{yaowang@g.clemson.edu}) and M.M. (\href{mailto:mmitrano@g.harvard.edu}{mmitrano@g.harvard.edu}) 
]{}
\affiliation{Department of Physics, Harvard University, Cambridge, Massachusetts 02138, USA}
\date{\today}

\begin{abstract}
\hspace{5.7cm}\textbf{ABSTRACT}\\
Manipulating spin fluctuations with ultrafast laser pulses is a promising route to dynamically control collective phenomena in strongly correlated materials. However, understanding how photoexcited spin degrees of freedom evolve at a microscopic level requires a momentum- and energy-resolved characterization of their nonequilibrium dynamics. Here, we study the photoinduced dynamics of finite-momentum spin excitations in two-dimensional Mott insulators on a square lattice. By calculating the time-resolved resonant inelastic x-ray scattering cross-section, we show that an ultrafast pump above the Mott gap induces a prompt softening of the spin excitation energy, compatible with a transient renormalization of the exchange interaction. While spin fluctuations in a hole-doped system (paramagnons) are well described by Floquet theory, magnons at half filling are found to deviate from this picture. Furthermore, we show that the paramagnon softening is accompanied by an ultrafast suppression of $d$-wave pairing correlations, indicating a link between the transient spin excitation dynamics and superconducting pairing far from equilibrium.

\end{abstract}
\maketitle

\section{Introduction}
Following the early demonstration of ultrafast demagnetization in ferromagnets\,\cite{Beaurepaire1996ultrafast}, the optical manipulation of spin degrees of freedom emerged as an effective strategy for steering electronic properties in quantum materials\,\cite{kirilyuk2010ultrafast}. Over the years, light excitation protocols evolved from purely thermal effects to more sophisticated optical excitations involving Zeeman interaction\,\cite{kampfrath2011coherent}, nonlinear phonon excitation\,\cite{nova2016effective,disa2020polarizing}, or the renormalization of local interactions\,\cite{Hendry2019,ron2020ultrafast}. While most of these experimental efforts aim to manipulate magnetic orders, driving fluctuating spins with light could represent an effective dynamical control strategy for superconductivity in strongly-correlated electron systems.  

In the copper oxides, experimental \cite{rossat1991neutron,mook1993polarized,fong1999neutron} and theoretical works \cite{scalapino1986d,gros1987superconducting,kotliar1988superexchange, schrieffer1989dynamic, scalapino1995case,tsuei2000pairing,maier2006structure} suggest that spin fluctuations around the antiferromagnetic wavevector $\Qbf_{\mathrm{AF}}=(\pi,\pi)$ could contribute to the pairing interaction and determine the superconducting critical temperature $T_c$. Whether optically driving these spin fluctuations may explain the recent observation of light-induced superconductivity in certain copper oxides\,\cite{fausti2011light,hu2014optically,kaiser2014optically,Nicoletti2014optically} or be conducive to further optimized light-induced coherence is an open question.

\begin{figure}[!t]
\begin{center}
\includegraphics[width=\columnwidth]{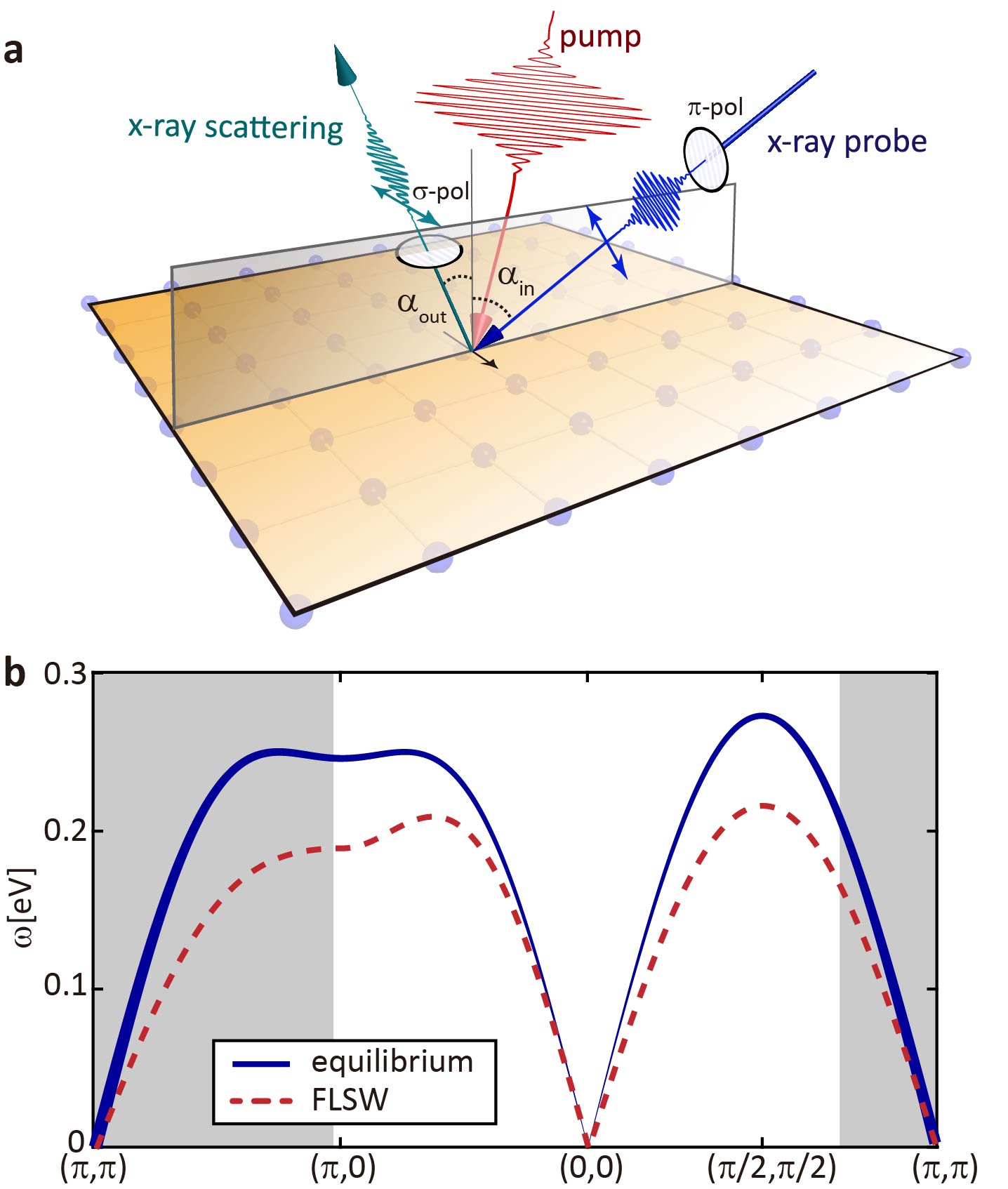}
\caption{\label{fig:cartoon} {\bf Probing light-driven spin fluctuations with resonant inelastic x-ray scattering. } \textbf{a} Sketch of a pump-probe resonant inelastic x-ray scattering (RIXS) experiment with $\pi$-$\sigma$ polarization. \textbf{b} Magnon dispersion in the 2D Hubbard model in (blue solid line) and out of equilibrium (red dashed line, pump field amplitude $A_0=0.6$) as determined through linear spin wave and Floquet linear spin wave (FLSW) theory [see Eqs.~ \eqref{eq:LSW}]. The gray shaded region indicates momenta  which are kinematically inaccessible to Cu $L$-edge RIXS.\vspace{-0.7cm}
}
\end{center}
\end{figure}

Directly probing the effect of ultrafast light pulses on spin fluctuations at finite momentum $\qbf$ requires a tool beyond optical probes, which are only sensitive to long-wavelength ($\qbf\sim0$) dynamics\,\cite{zhao2011coherent,dal2012disentangling,batignani2015probing,yang2020ultrafast}. Resonant inelastic x-ray scattering (RIXS) at transition-metal $L$-edge RIXS with cross-polarized incident and scattered x-rays (see Fig.~\ref{fig:cartoon}\textbf{a}) is sensitive to magnetic excitations, and is able to map their dispersion thanks to the high photon momentum \,\cite{ament2009theoretical,haverkort2010theory, ament2011resonant}. Pioneering time-resolved resonant inelastic x-ray scattering (trRIXS) experiments in the hard x-ray regime provided the first snapshots of pseudospin dynamics near $\Qbf_{\mathrm{AF}}$ in layered iridates~\cite{dean2016ultrafast,mazzone2020trapped} and future soft X-ray experiments at the Cu and Ni $L$-edges will further expand the reach of this new ultrafast technique.  Based on these developments, it becomes critical to theoretically, and possibly experimentally, investigate how dispersive spin fluctuations and superconducting pairing evolve in a prototypical correlated system driven far from equilibrium. 

In this work, we present a trRIXS calculation of finite-momentum spin fluctuations in the paradigmatic single-band Hubbard model. We calculate the full trRIXS cross section for a direct $L_3$-edge absorption for $\pi$-$\sigma$ polarizations and for two different hole doping levels. We show that an ultrafast resonant pump induces a prompt softening of the short-ranged spin excitation spectrum, compatible with a light-induced renormalization of the exchange interaction [sketched in Fig.~\ref{fig:cartoon}\textbf{b}], and a simultaneous suppression of $d$-wave pairing correlations.
Our findings in a canonical Hubbard model demonstrate that light pulses can be used to tailor finite-momentum spin fluctuations and pave the way to novel strategies for the light control of unconventional superconductors.

\section{Results and Discussion}
\subsection{Microscopic model}
With a particular focus on the high-$T_c$ cuprates, we consider valence and conduction electrons described by the 2D single-band Hubbard model and for two different hole dopings ($x=0$ and $x=0.167$). Since the x-ray scattering explicitly involves an intermediate core level, the full Hamiltonian reads as 
\begin{eqnarray}\label{eq:Hubbard}
	\Ham &=&  -\sum_{\ibf,\jbf,\sigma}\!t_h^{\ibf\jbf} ( d_{\jbf\sigma}^\dagger d_{\ibf\sigma}\!+\!h.c.)\! + U \sum_{i} n^{(d)}_{\ibf\uparrow}n^{(d)}_{\ibf\downarrow}\nonumber\\
	&&+ \sum_{\ibf\alpha\sigma} E_{\rm edge}(1-n^{(p)}_{\ibf\alpha\sigma}) - U_c \sum_{\ibf\alpha\sigma\sigma^\prime} n^{(d)}_{\ibf\sigma}(1-n^{(p)}_{\ibf\alpha\sigma^\prime})\nonumber\\ &&+\lambda\sum_{i\alpha\alpha^\prime\atop \sigma\sigma^\prime}p_{\ibf\alpha\sigma}^\dagger \chi_{\alpha\alpha^\prime}^{\sigma\sigma^\prime}p_{\ibf\alpha^\prime\sigma^\prime}.
\end{eqnarray}
Here, $d_{\ibf \sigma}$ ($d_{\ibf \sigma}^\dagger$) annihilates (creates) a $3d_{x^2-y^2}$ electron at site $\ibf$ with spin $\sigma$, while $p_{\ibf \alpha \sigma}$ ($p_{\ibf \alpha \sigma}^\dagger$) annihilates (creates) a $2p_{\alpha}$ electron ($\alpha\! =\! x, y, z$). $t_h^{\ibf\jbf}$ is the hopping integral between site $i$ and $j$, $n^{(d)}_{\ibf}\!=\!\sum_\sigma d_{\ibf \sigma}^\dagger d_{\ibf \sigma}$ and $n^{(p)}_{\ibf\alpha\sigma}\!=\! p_{\ibf\alpha \sigma}^\dagger p_{\ibf\alpha \sigma}$ are the valence and core-level electron density operators, respectively. The nearest-neighbor hopping amplitude is set to $t_h=300$\,meV, the next-nearest neighbor hopping to $t^\prime_h \!=\!-0.3\,t_h$, and the on-site repulsion to $U\!=\!8t_h$ \cite{jia2014persistent,jia2016using}. The core-hole potential $U_c$ is instead fixed at 4$t_h$ and regarded identical for all 2$p$ orbitals\,\cite{tsutsui2000resonant,jia2014persistent,jia2016using}. $E_{\rm edge}=938$\,eV represents instead the Cu $L$-edge absorption energy, i.e., the energy difference between the $3d$ and $2p$ orbitals without spin-orbit coupling. Finally, the spin-orbit coupling with the x-ray-induced core hole in the degenerate $2p$-orbitals is accounted by the last term in Eq.~\eqref{eq:Hubbard} with $\lambda=13$\,eV\,\cite{tsutsui2000resonant,kourtis2012exact}.

\begin{figure}[!t]
\begin{center}
\includegraphics[width=\columnwidth]{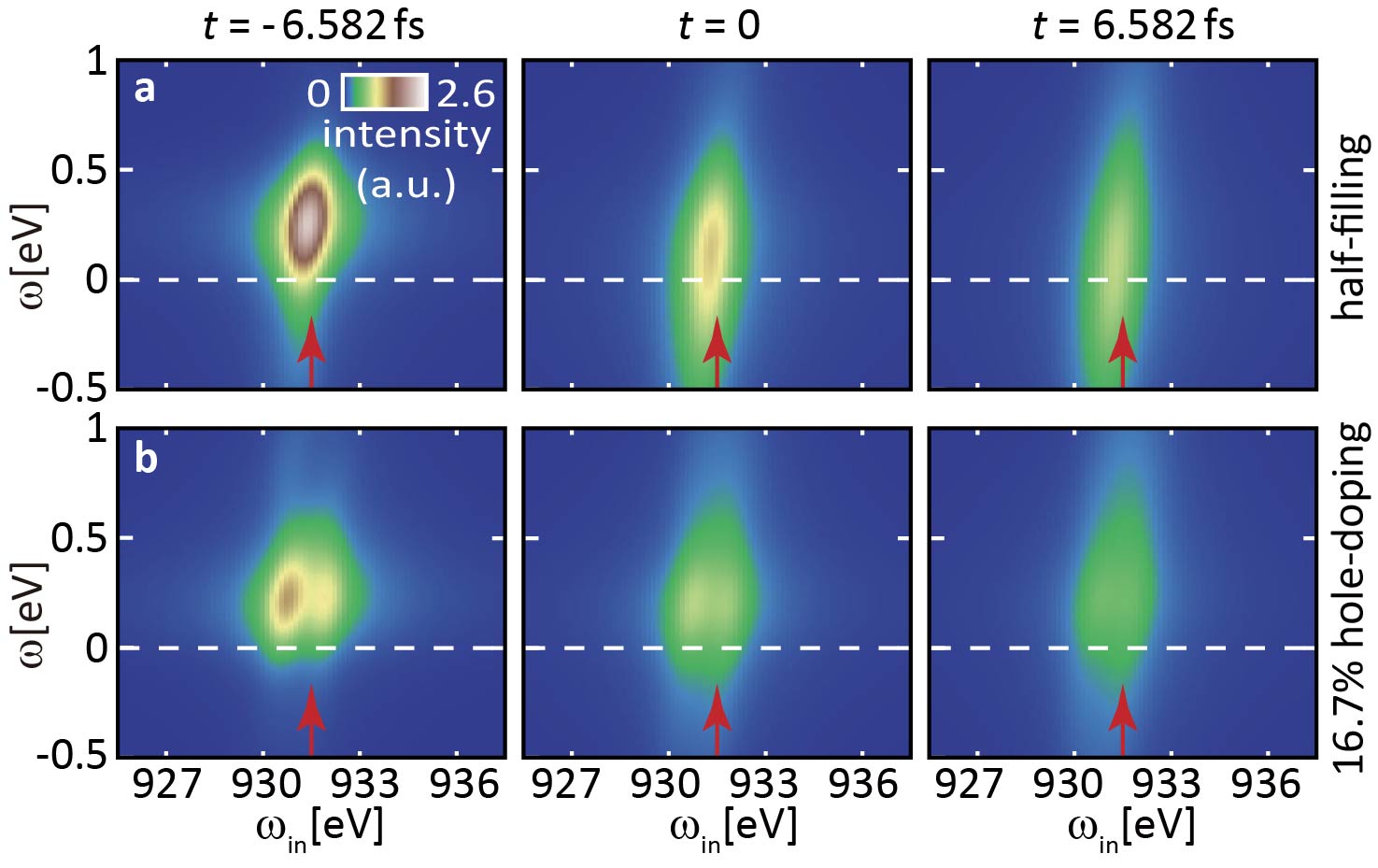}
\caption{\label{fig:spectra} {\bf Snapshots of the time-resolved resonant inelastic x-ray scattering signal.} Cross-polarized time-resolved resonant inelastic x-ray scattering (trRIXS) spectra for selected pump-probe time delays as function of the incident photon energy $\win$ for \textbf{a} $x=0$ and \textbf{b} $x=0.167$ hole doping. $\omega=\win-\wout$ indicates the energy loss, while the momentum transfer is kept fixed at $\qbf=(\pi/2,\pi/2)$. The peak at $\omega\sim0.3$ eV corresponds to  the collective spin fluctuations of the system. The red arrow marks the center of mass of the resonance along the $\win$ axis.  The color bar indicates the spectral intensity in arbitrary units (a.u.).
}
\end{center}
\end{figure}

The trRIXS cross-section at a direct absorption edge can be written as \cite{chen2019theory, wang2020time} (see Supplementary note 1 for further details)
\begin{eqnarray}\label{eq:RIXScrossSec}
\mathcal{I}(\qbf,\wout,\win,t)\!&\mkern-6mu=&\!\mkern-12mu\iiiint
e^{i\win(t_2-t_1)-i\wout(t_2^\prime-t_1^\prime)} 
  g(t_1;t)g(t_2;t) \nonumber\\
&&\mkern-42mu\times \big\langle \hat{\mathcal{D}}_{\qin\epsin}^\dagger(t_2)\hat{\mathcal{D}}_{\qout\epsout}(t_2^\prime) \hat{\mathcal{D}}_{\qout\epsout}^\dagger(t_1^\prime)   \hat{\mathcal{D}}_{\qin\epsin}(t_1)  \big\rangle\nonumber\\
&&\mkern-42mu\times l(t_1^\prime-t_1)l(t_2^\prime-t_2) dt_1  dt_2 dt_2^\prime dt_1^\prime\,,  
\end{eqnarray}
where ${\bf q}=\qin-\qout$ is the momentum transfer between incident and scattered photons with energies $\win$ and $\wout$ respectively, and $l(\tau)\! =\!  e^{- \tau/\tau_{\rm core}}\theta(\tau)$ the core-hole lifetime decay, with $\tau_{\rm core}\! =\! 0.5\,t_h^{-1}$ as its characteristic timescale.
For a direct transition (e.g. Cu $L$-edge), the dipole operator is explicitly given by $\mathcal{D}_{\qbf\eps}=\sum_{\ibf\alpha\sigma}e^{-i\qbf\cdot \rbf_\ibf}(A_{\alpha\eps} d_{\ibf\sigma}^\dagger p_{\ibf\alpha\sigma}  + h.c.)$
where $A_{\alpha\eps}$ is the matrix element of the dipole transition associated with a $\eps$-polarized photon.
As shown in Fig.~\ref{fig:cartoon}\textbf{a}, we employed the $\pi-\sigma$ polarization configuration ($\epsin$ parallel and $\epsout$ perpendicular to the scattering plane) and fixed $\alpha_{\rm in}+ \alpha_{\rm out} = 50^{\circ}$. This probe condition maximizes the spin-flip response in presence of the spin-orbit coupling term\,\cite{ament2009theoretical,jia2014persistent}. Calculations for the $\pi$-$\pi$ polarization configuration are reported in Supplementary note 2 for completeness. Due to the $O(N_t^4)$ complexity of the trRIXS calculation, we adopt the 12D Betts cluster as a compromise between complexity and finite-size.

\begin{figure*}[!t]
\begin{center}
\includegraphics[width=18cm]{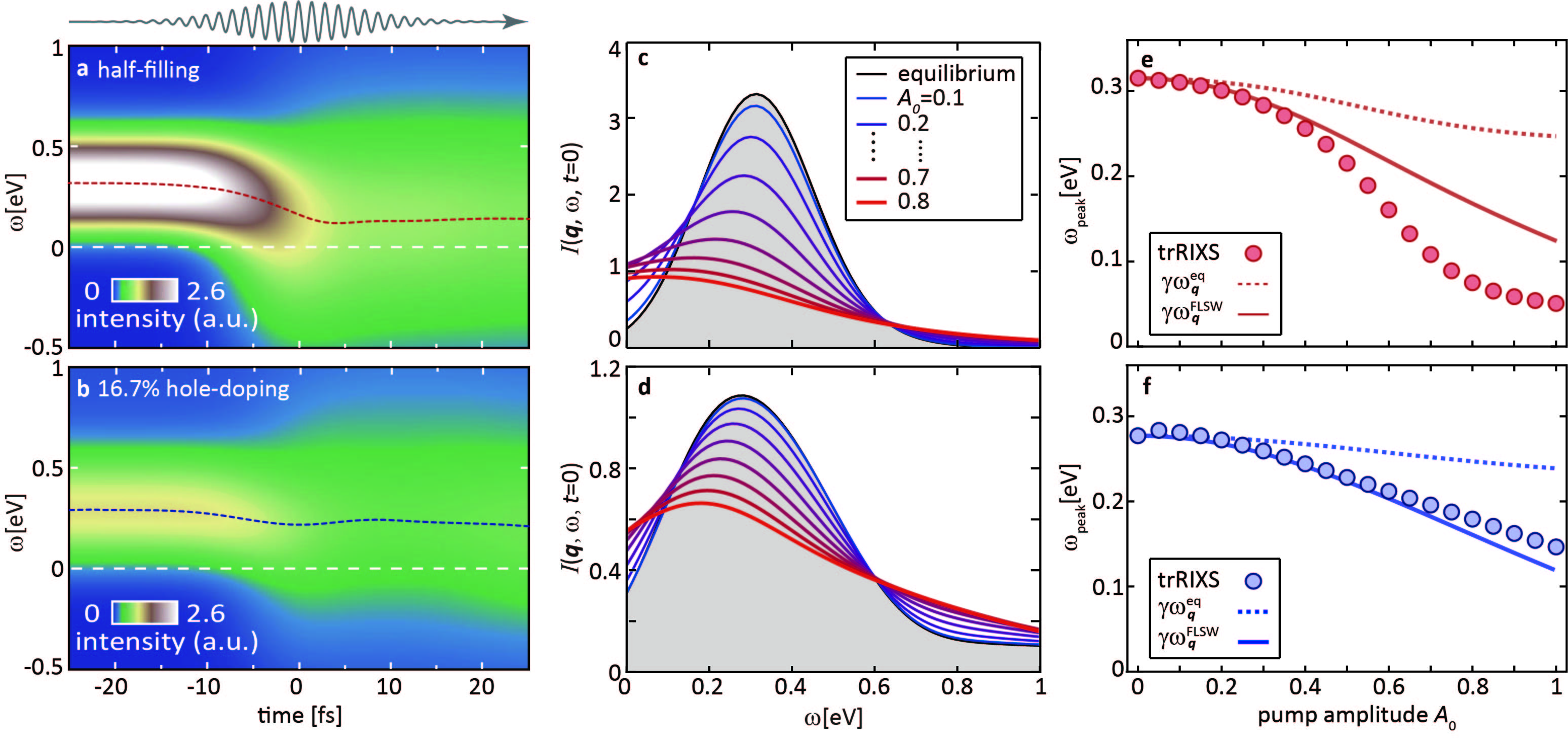}
\caption{\label{fig:comparison}{\bf Time-dependent renormalization of the spin fluctuations.}
\textbf{a,b} Time evolution of the cross-polarized time-resolved resonant inelastic x-ray scattering (trRIXS) spectra for pump energy $\Omega=10t_h$ and amplitude $A_0=0.6$ at \textbf{a} $x=0$ and \textbf{b} $x=0.167$ doping. The momentum transfer is fixed at $\qbf=(\pi/2,\pi/2)$. The thin dashed line tracks the instantaneous peak positions of the spin fluctuation peak. The color bar indicates the spectral intensity in arbitrary units (a.u.). trRIXS spectra at $t=0$ (center of the pump pulse) for variable pump amplitudes at \textbf{c} $x=0$ and \textbf{d} $x=0.167$ doping. Spin excitation energy $\omega_{{\bf q}}$ (the peak of the trRIXS intensity) as function of the pump amplitude at  \textbf{e} $x=0$ and \textbf{f} $x=0.167$ doping. The energy width of these spectra includes a broadening due to the finite x-ray probe duration. The dashed line denotes pure photodoping effects ($\gamma\omega^{\mathrm{eq}}_{\bf q}$, where $\omega^{\mathrm{eq}}_{\bf q}$ is the numerically evaluated magnon energy at equilibrium and $\gamma$ is the instantaneous local moment defined in the text), while the solid line represents a Floquet linear spin wave prediction of the magnon energies (including photodoping effects) ($\gamma\omega^{\mathrm{FLSW}}_{\bf q}$).}

\end{center}
\end{figure*}

In our calculations, the vector potential of the optical pump field is 
\begin{equation}
    \mathbf{A}^{\rm(pump)}(t) = A_0\, e^{-t^2/2\sigma_t^2} \cos(\Omega t)\, \hat{\mathbf{e}}_{\rm pol},
\end{equation}
with frequency $\Omega=10\,t_h\sim3.0$ eV (above the Mott gap $\sim\! 4-5\,t_h$) and duration $6\sigma_t=18\,t_h^{-1}=39.49$\,fs, while the probe has a Gaussian envelope $g(\tau;t) =  e^{-(\tau-t)^2/2\sigma_{\rm pr}^2}/{\sqrt{2\pi}\sigma_{\rm pr}}$ with $\sigma_{\rm pr}\! =\! 1.5\,t_h^{-1}$. $A_0$ is measured in natural units $ea_0E_0/\Omega$, where $A_0=0.1$ corresponds to a $E_0=79$\,mV/$\mathrm{\AA}$ peak electric field.
The pump polarization $\hat{\mathbf{e}}_{\rm pol}$ is linear along the $x$ direction. 
\par Here, we numerically calculate the trRIXS cross section at each time step $t$ by scanning the full four-dimensional hyperplane defined by the time-integration variables in eq. (\ref{eq:RIXScrossSec}) at zero temperature.
We first determine the equilibrium ground-state wavefunction with the parallel Arnoldi method\,\cite{lehoucq1998arpack,  jia2017paradeisos} and, then, calculate the time evolution of the system through the Krylov subspace technique\,\cite{manmana2007strongly, balzer2011krylov}.

\subsection{trRIXS and transient spin fluctuations}
In Figure \ref{fig:spectra}, we show selected snapshots of the cross-polarized trRIXS spectrum at $\qbf=(\pi/2,\pi/2)$, where the magnon dispersion is maximized [see Supplementary note 3 for a comparison with $\qbf=(0,2\pi/3)$). At resonance ($\win\sim 931$ eV for Cu  $L_3$-edge in Fig.~\ref{fig:spectra}], the loss spectra below the Mott gap are dominated by a single intense peak at $\omega\sim2J=0.3$ eV, where $J=4t_h^2/U$ is the exchange interaction. The energy and shape of this peak reflect the distribution of spin excitations: for an undoped Hubbard model, these manifest as a coherent magnon due to the strong antiferromagnetic (AFM) order; in contrast, in the $x=0.167$ case, the spin excitations damp into a short-ranged paramagnon~\cite{le2011intense,dean2013persistence, dean2013high, lee2014asymmetry, ishii2014high}. At negative time delays, when the system is close to equilibrium, our calculations reveal a splitting of the resonance along the $\win$ axis at $x=0.167$ doping. This reflects the presence of two different chemical environments, as the valence states can be either half-occupied or empty.

At the center of the pump pulse, the resonance undergoes (i) an overall suppression, and (ii) a redshift along the energy loss axis for both compositions. 
Since the center of the absorption peak does not exhibit a pump-induced shift along $\win$, one can fix $\win=931.5$\,eV and examine the time evolution of the magnon/paramagnon peak at ${\bf q}=(\pi/2,\pi/2)$ [see Figs.~\ref{fig:comparison}\textbf{a} and \textbf{b}]. The intensity drop and the peak shift closely follow the pump envelope, and then persist long after the pump arrival due to the lack of dissipation. The photoexcited RIXS spectrum is visibly broadened at $x=0$ and to a lesser extent at $x=0.167$. 
Next, we analyze the dependence of the spin fluctuation peak as function of the pump field strength $A_0$ up to a maximum value of 1. This physical regime is relevant to recent experiments on Sr$_2$IrO$_4$ \cite{dean2016ultrafast} and Sr$_3$Ir$_2$O$_7$ \cite{mazzone2020trapped}, which employed fields strengths $A_0$ up to 0.35 and 0.52, respectively.
As $A_0$ increases, the peak position for both $x=0$ and $x=0.167$ shifts monotonically to lower energies [see Figs.~\ref{fig:comparison}\textbf{c} and \textbf{d}] by $~80\%$ and $~50\%$, respectively. At the same time, the magnon/paramagnon peak broadens due to the generation of high-order fluctuations beyond simple spin-wave excitations and of photo-doped carriers which perturb the spin background. These observations imply that optical laser pulses with negligible momentum have tangible effects on the spin fluctuations at large momenta. 

\subsection{Magnon/paramagnon energy renormalization}
While we report here an exact calculation of the full trRIXS cross-section, it is useful to discuss the microscopic physics at play in more intuitive terms. The magnon/paramagnon energy softening in our single-band Hubbard model can be attributed to photodoping, a dynamical Floquet-type renormalization of the spin exchange interaction $J$\,\cite{mentink2014ultrafast,mentink2015ultrafast}, or a combination of the two. Other experimentally relevant mechanisms, such as magnetophononics \cite{Fechner2018magnetophononics, nova2016effective,Afanasiev2021ultrafast}, would require extending the Hubbard Hamiltonian through the inclusion of additional degrees of freedom. 

We first consider the photodoping contribution. Due to the light-induced holon-doublon excitations, the local spin moment $\langle m_z^2\rangle$ is diluted. In the simplest linear spin wave picture, this will induce a homogeneous softening of the spin structure factors across the entire Brillouin zone. To quantify this effect, we calculate the instantaneous local moment at $t=0$, defined as $\gamma = \sqrt{\frac{\langle \mathbf{S}^2(t=0)\rangle}{\langle \mathbf{S}^2(t=-\infty)\rangle}}\,$, and plot the renormalized magnon/paramagnon peak energy $\gamma \omega_{\qbf}^{\rm eq}$ as dashed lines in Fig.~\ref{fig:comparison}{\bf e}-{\bf f}. Here, $\omega_{\qbf}^{\rm eq}$ is the numerically-evaluated magnon energy at equilibrium. As visible from the comparison, the local photodoping effect, i.e. the dilution of the local spin moment, is too small to account for the observed energy softening.

\par The other possible mechanism to explain the peak shift hinges on a light-induced renormalization of the effective Hamiltonian. It has been demonstrated that, for an ideal system driven by an infinitely long, nonresonant pump, the effective spin exchange energy $J$ is modified by a Floquet dressing of the intermediate doubly-occupied states\,\cite{mentink2015ultrafast}. In order to quantitatively understand our trRIXS spectrum,  we apply Floquet linear spin wave (FLSW) theory at different pump field amplitudes. In this framework, we assume the system at the center of the pump pulse to be in a steady-state, thus leading to a closed form for the spin excitations dispersion $\omega_\qbf^{\rm FLSW}$ (see Methods section.)

By including both the Floquet renormalization and photodoping effects, we calculate the transient magnon/paramagnon energy $\gamma\omega_{\qbf=(\pi/2,\pi/2)}^{\rm FLSW}$ and multiply it by a constant factor to match the equilibrium peak positions (see Fig.~\ref{fig:comparison}{\bf e}-{\bf f}). The scale factors are 1.15 for $x=0$ and 1.02 for $x=0.167$. At $x=0$, the FLSW prediction tracks the magnon softening for weak pump fields but grossly deviates at higher pump strength. This is likely due to mobile photo-carriers which frustrate the AFM ground state~\cite{tsutsui2020antiphase} and lower the energy cost of spin flip events. At $x=0.167$ hole-doping, however, FLSW theory closely aligns with the trRIXS calculation for all pump strengths. 
We tentatively attribute the difference between these two trends to intrinsic three-site correlated hopping. In the low-energy projection of the Hubbard Hamiltonian, holes can hop to next- and next-next-nearest neighbors (and thus lower the total energy) provided that the middle site has an opposite spin configuration\,\cite{spalek1988effect,stephan1992single,bal1995spin}, which is otherwise inhibited. At equilibrium, this term partially cancels the doping-induced energy softening\,\cite{jia2014persistent,parschke2019numerical} and is essential to reproduce the spin excitation spectrum of copper oxides.
Active at $x=0.167$, these correlations are silent in the AFM ground state and build up too slowly to compensate the spin frustration. While both Floquet and linear spin wave theories over-simplify the nonequilibrium many-body physics at play, their agreement with the paramagnon energy shift provides evidence of an ultrafast modification of effective model parameters beyond pure photodoping.

\begin{figure}[!t]
\begin{center}
\includegraphics[width=\columnwidth]{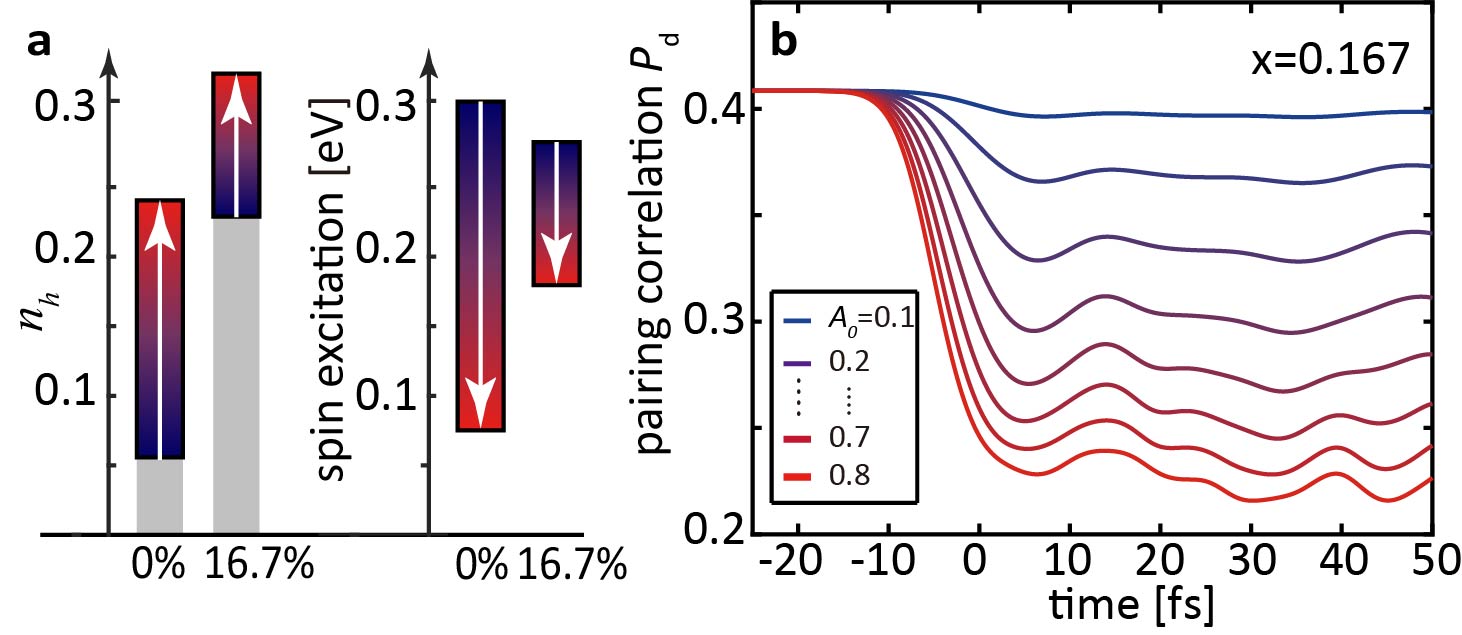}
\caption{\label{fig:pairing} {\bf Pump-induced modification of spin excitation energy and pairing correlations.} \textbf{a} Equilibrium (gray) and maximal transient (colored) change of hole concentration and spin excitation energy (extracted from time-resolved resonant inelastic x-ray scattering spectral peaks) for the pump conditions in panel Fig.~\ref{fig:comparison} at half-filling and 16.7\% hole doping, respectively. \textbf{b} Corresponding dynamics of the $d$-wave pairing correlations. 
}
\end{center}
\end{figure}

This striking light-induced softening represents a genuine emergent phenomenon, which is quantitatively distinct from the behavior of magnons/paramagnons at equilibrium. Short-range spin excitations in La$_2$CuO$_4$ show negligible dispersion changes for temperature changes well above their N\'{e}el temperature\,\cite{Matsuura2017development}, while hole-doped cuprates exhibit at most a redshift of order $20-30\%$ for a doping change $\Delta x=0.4$ (undoped to overdoped regime)\,\cite{le2011intense,dean2013persistence, dean2013high, lee2014asymmetry, ishii2014high}. For a direct comparison, we compute the transient hole concentrations $\langle n_h(t=0)\rangle$ [see Figs.~\ref{fig:pairing}\textbf{a} and Supplementary Note 4], defined as $n_h = \sum_\ibf (1-n_{\ibf\uparrow}) (1-n_{\ibf\downarrow})/N$. Due to the presence of quantum fluctuations, the $\langle n_h\rangle$ at equilibrium is 0.06 higher than the nominal hole concentration (0.167). Out of equilibrium at $t=0$, the strongest pump gives only an extra 0.07 holes per site. In both the doped and undoped systems, the number of photoinduced holes is less than the full doping excursion between underdoped and overdoped regimes, and yet produces a larger softening than what is observed in cuprates at equilibrium.

\subsection{$d$-wave pairing correlations}
Since spin fluctuations have been discussed as a possible source of pairing in the cuprates\,\cite{ scalapino1986d,gros1987superconducting,kotliar1988superexchange, schrieffer1989dynamic, scalapino1995case,tsuei2000pairing,maier2006structure,scalapino2012common}, we examine whether the paramagnon energy renormalization affects the superconducting pairing.
We calculate the $d$-wave pairing correlation function $P_d = \langle \Delta_d^\dagger \Delta_d \rangle$ at $x=0.167$ for various pump strengths [see Fig.~\ref{fig:pairing}\textbf{b}], with the factor $\Delta_d$ defined as $\Delta_d = \frac1{\sqrt{N}} \sum_\kbf d_{\kbf\uparrow} d_{-\kbf\downarrow} \left[\cos k_x - \cos k_y\right]$. At the pump arrival, $P_d$ undergoes a rapid suppression, thus indicating a decrease of pairing in the $d$-wave channel. The drop in $P_d$ is found to be monotonic with the increase of the pump strength $A_0$. Intuitively, the value of $P_d$ is determined by both the quasiparticle density and the pairing interaction strength. In our calculations, the pump simultaneously increases quasiparticles via photodoping and reduces the spin excitation energy. While the former effect is favorable for superconductivity, it is the latter that dominates and reduces the nonequilibrium pairing correlations.

\section{Conclusions}

We studied the light-induced dynamics of finite-momentum spin excitations in the prototypical 2D Hubbard model through a full calculation of the $\pi$-$\sigma$ polarized trRIXS spectrum. By exciting a Mott insulator with an ultrashort pump pulse, we find that magnons and paramagnons undergo a dramatic light-induced softening, which cannot be exclusively explained by photodoping. Paramagnon excitations are quantitatively described by a Floquet renormalization of the effective exchange interactions, while magnons at high field strengths deviate from this picture. At optimal doping, we also observe a sizeable reduction of $d$-wave pairing correlations. While here we explore a paradigmatic optical excitation at the Mott gap, alternative mechanisms such as phonon Floquet\,\cite{hubener2018phonon}, ligand manipulation\,\cite{ron2020ultrafast}, multi-band dynamical effects\,\cite{tancogne2018ultrafast,golevz2019multiband}. and coupling to cavities\,\cite{Gao2020photoinduced, Sentef2020quantum} might lead to harder spin fluctuations and increased pairing \cite{mentink2015ultrafast,coulthard2017enhancement}. Future polarization-dependent trRIXS experiments and theoretical calculations will be key to investigate these photoinduced spin dynamics in a wide variety of quantum materials \cite{cao2019ultrafast,mitrano2020probing}.

\section{Methods}
\subsection{Floquet Linear Spin Wave Theory}
To describe the evolution of magnetic excitations at finite momentum, we generalize the local Floquet renormalization of the spin-exchange interaction\,\cite{mentink2014ultrafast,mentink2015ultrafast} through Floquet linear spin wave theory. This approach is based on the assumption that the driven system can be instantaneously approximated by an effective Hamiltonian $\Ham_{\rm F}$ and a steady state wavefunction $|\psi_F\rangle$. In the Floquet framework, the hopping $t_{ij}$ in the effective Hamiltonian $\Ham_{\rm F}$ is renormalized by a factor $\mathcal{J}_m(\mathbf{A}\cdot \mathbf{r}_{ij})$, where $\mathcal{J}_m(x)$ is the Bessel function of the first kind and $\mathbf{A}$ is the vector potential of an infinitely long driving field. Following second-order perturbation theory, the effective spin-exchange interaction between two coordinates is renormalized into~\cite{mentink2015ultrafast}
\begin{eqnarray}
J_{ij}^{\rm eff} = \frac{4t_{ij}^2}{U} \sum_{m=-\infty}^{\infty} \frac{\mathcal{J}_m(\mathbf{A}\cdot \mathbf{r}_{ij})^2}{1+m\Omega/U}\,.
\end{eqnarray}
Then, through the standard Holstein–Primakoff transformation and large-$S$ expansion, one can obtain the linearized bosonic Hamiltonian
\begin{eqnarray}
\Ham_{FLSW} = \sum_{\qbf} \sqrt{A_\qbf^2 - B_\qbf^2}\,a_\qbf^\dagger a_\qbf = \sum_{\qbf}\omega_\qbf^{\rm FLSW} a_\qbf^\dagger a_\qbf\,,
\end{eqnarray}
where $a_\qbf$ is the magnon annihilation operator and the coefficients $A_\qbf$ and $B_\qbf$ in the two-dimensional plane are~\cite{delannoy2009low}
\begin{eqnarray}
A_\qbf &=& J_x^{\rm eff} + J_y^{\rm eff} - J_{xy}^{\rm eff} \left[1-\cos (q_x+ q_y)\right]\nonumber\\
&&- J_{yx}^{\rm eff} \left[1-\cos (q_x- q_y)\right]\,,\\
B_\qbf &=& J_x^{\rm eff}\cos q_x + J_y^{\rm eff}\cos q_y\,.
\label{eq:LSW}
\end{eqnarray}
Here, unlike the equilibrium counterpart, the coefficients are anisotropic due to the electric field polarization:
\begin{eqnarray}
J_{x/y}^{\rm eff} &=& \frac{4t_h^2}{U} \sum_{m=-\infty}^{\infty} \frac{\mathcal{J}_m(\mathbf{A} \cdot \mathbf{e}_{x/y})^2}{1+m\Omega/U}\,,\\
J_{xy/yx}^{\rm eff} &=& \frac{4t_h^{\prime2}}{U} \sum_{m=-\infty}^{\infty} \frac{\mathcal{J}_m(\mathbf{A}\cdot (\mathbf{e}_{x} \pm \mathbf{e}_{y}))^2}{1+m\Omega/U}\,.
\end{eqnarray}
This derivation can be extended to higher order following the strategy of equilibrium LSW theory\,\cite{delannoy2009low}.

\section*{Data Availability}
The numerical data that support the findings of this study are available from the corresponding authors upon reasonable request.

\section*{Code Availability}
The relevant scripts of this study are available from the corresponding authors upon reasonable request.

\section*{Acknowledgements} 
We thank D.R. Baykusheva, M. Buzzi, C.-C. Chen, M. P. M. Dean, A. A. Husain, C. J. Jia, D. Nicoletti, and M. Sentef for insightful discussions.  Y.C., T.P.D., and B.M. were supported by the US Department of Energy, Office of Basic Energy Sciences, Division of Materials Sciences and Engineering, under Contract No. DE-AC02-76SF00515. M. M. was supported by the William F. Milton Fund at Harvard University. This research used resources of the National Energy Research Scientific Computing Center (NERSC), a U.S.~Department of Energy Office of Science User Facility operated under Contract No.~DE-AC02-05CH11231.

\section*{Author contributions} 
Y.W. and M.M. conceived the project. Y.W. and Y.C. performed the calculations. Y.W. and M.M. performed the data analysis. Y.W. and M.M. wrote the paper with the help from Y.C., B.M., and T.P.D.

\section*{Competing interests} The authors declare no competing interests.

\section*{References}
\bibliography{papers}
\end{document}